\documentclass[sigconf]{acmart}
\AtBeginDocument{%
  }
  
\usepackage{times}
\usepackage{enumitem}
\usepackage{graphicx}

\usepackage{mdframed}
\usepackage{placeins}  

\usepackage{xspace}
\usepackage{hyperref}
\usepackage{url}
\usepackage{algorithm}
\usepackage{algorithmicx}
\usepackage{algpseudocode}
\usepackage{graphicx}
\usepackage{subcaption}
\usepackage{booktabs}
\usepackage{comment}
\usepackage{enumitem}
\usepackage{pgfplots}
\usepackage{tabularx}
\usepackage{multirow}
\usepackage{svg}
\usepackage{tcolorbox}
\usepackage{wasysym}

\def\sys{Privy\xspace}

\title{Privy: From Fine Print to Fair Practice in Privacy Rights Exercise}

\author{Qi Sun}
\affiliation{%
  \institution{Johns Hopkins University}
  \city{Baltimore}
  \state{MD}
  \country{USA}}
\email{qsun28@jh.edu}

\author{Ziyang Li}
\affiliation{%
  \institution{Johns Hopkins University}
  \city{Baltimore}
  \state{MD}
  \country{USA}}
\email{ziyang@cs.jhu.edu}

\author{Yinzhi Cao}
\affiliation{%
  \institution{Johns Hopkins University}
  \city{Baltimore}
  \state{MD}
  \country{USA}}
\email{yinzhi.cao@jhu.edu}

\author{Yaxing Yao}
\affiliation{%
  \institution{Johns Hopkins University}
  \city{Baltimore}
  \state{MD}
  \country{USA}}
\email{yaxing@jhu.edu}

\begin{document}

\begin{abstract}
Privacy regulations such as the CCPA and GDPR grant individuals rights over their personal data, yet it remains challenging for most users to exercise them in practice due to vague policy interpretation and unapproachable settings on web interfaces. We introduce Privy, an LLM-powered browser assistant that guides users through exercising their privacy rights on websites. Privy automatically analyzes a website's privacy policy and surfaces the specific rights available as action labels in a side panel. When a user selects a right, Privy provides step-by-step guidance and navigation, presenting direct links, generating email templates, or guiding form completion. Users can also request on-demand policy evidence and rights education to enhance their literacy. 
A technical evaluation across 14 websites shows that Privy extracts rights with high precision (0.979) and completes 96.3\% of privacy tasks in an average of 3.2 steps. A user study ($N=15$) also demonstrates the overall high-level of perceived helpfulness among users. Our findings suggest that comprehension and usability are not two separate challenges but a single interaction problem, and that effective privacy support requires integration of policy understanding and privacy actions. We offer design suggestions for future privacy assistants. 
\end{abstract}

\maketitle


\begin{figure}[!t]
  \centering
  \includegraphics[width=\linewidth]{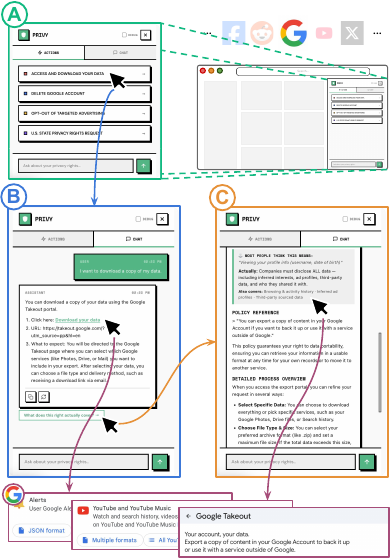} 
  \caption{User Interface of \sys. \sys serves as a side panel of the webpage, which consists of three main components: (A) \textbf{Action labels} extracted from the privacy policy surface available  rights as selectable options; (B) After selecting a right, \sys provides step-by-step guidance to the relevant service (here, \sys provides a direct link to Google Takeout based on the Google's privacy policy); (C) Users can request on-demand policy context, including legal references, verbatim policy excerpts, and rights education.} \label{fig:interface} 
\end{figure}

\section{Introduction}

Modern privacy regulations, such as the GDPR, CCPA, and PIPL, have granted billions of individuals legal rights over their personal data, including the rights to access, delete, correct, and opt out of the sale or sharing of their information~\cite{acquisti2015privacy}.
These data subject rights (DSRs) represent a fundamental shift in the balance of power between individuals and organizations: for the first time, users can demand to know what data a company holds about them, request its deletion, or prohibit its sale to third parties.

Despite these legal guarantees, the vast majority of people have never exercised a single data subject right~\cite{draper2019digital}.
The reason is not a lack of legal protection but a persistent \textbf{policy--practice divide}~\cite{petelka2022generating,bowyer2022human}, i.e., the gap between what the law promises and what users can actually accomplish.
Consider a user who wants to stop a social media platform from selling her browsing data to advertisers.
She must first find and read a multi-thousand-word privacy policy to determine whether the platform even recognizes this right---a task most users skip entirely due to length, legal jargon, and inaccessibility~\cite{mcdonald2008cost}.
If the policy does mention an opt-out, it may describe the mechanism as ``adjust your privacy settings,'' ``visit our privacy choices page,'' or ``email privacy@example.com'', each requiring a completely different course of action. She must then locate the correct interface element on the website, which might be a toggle buried three menus deep, a link hidden in the footer, or a separate web form on a different page entirely~\cite{pins2022finding}. Even after completing the action, she has no way to confirm that it worked: the only feedback is an ambiguous ``Settings saved'' message, and she is left wondering whether she opted out of \textit{all} data sharing or merely adjusted a notification preference~\cite{habib2020scavenger}. Across organizations, these barriers compound: each website implements the same legal right through different policies, different interfaces, and different mechanisms, making every new attempt feel like starting from scratch~\cite{bowyer2022human,zimmeck2023usability}.

Exercising a DSR thus requires users to perform three sequential practices, each demanding knowledge and skills that most lack.
First, they must \textit{discover} which rights an organization recognizes and how each can be exercised---information that is encoded in privacy policies most users never read.
Second, they must \textit{navigate} the specific interface mechanisms the organization provides---settings pages, web forms, email addresses, or opt-out toggles---which vary widely across websites and are often deliberately obscured through dark patterns~\cite{lobel2024access}.
Third, they must \textit{understand} whether their actions were correct, sufficient, and will be honored---a judgment that requires knowledge of the legal scope of each right, which most users dramatically underestimate~\cite{kang2015mydata,turow2023consent}.
Without support for all three practices, users either abandon the process before starting, complete it without confidence, or exercise their rights only partially.

Prior research in HCI and usable privacy has made significant progress on the first two fronts, but in isolation.
Policy comprehension tools (e.g., structured privacy nutrition labels~\cite{kelley2009nutrition}, NLP-powered analyzers~\cite{fawaz2017polisis}, LLM-based conversational assistants~\cite{freiberger2025you}) help users understand what a policy says, but do not extract actionable rights or guide users through the exercise process.
On the other hand, DSR automation frameworks formalize submission workflows into replayable scripts~\cite{leschke2023streamlining} or transmit universal opt-out signals~\cite{zimmeck2023usability}, but these approaches are generic (e.g., pre-scripted for a fixed set of websites or operating at a protocol level invisible to the user) and they provide no transparency into what is being exercised or whether it succeeds.
In short, comprehension tools produce \textit{informed} users who remain stuck, and automation tools produce \textit{completed} requests that users do not understand.
No existing system bridges the full gap: contextualizing rights within organizational policies, guiding users through organization-specific interfaces, and building confidence through transparent evidence and education.

In this paper, we present \textbf{\sys}, an LLM-empowered browser assistant that guides the general public through the full process of exercising their privacy rights on any website.
\sys operates directly within the user's browsing environment as a Chrome extension with a persistent side panel, integrating three key capabilities:
(1)~\textbf{Policy Discovery and Analysis}: \sys automatically locates and analyzes the target website's privacy policy, extracting the specific rights the organization offers along with the concrete mechanism (email, link, navigation path, or form) for exercising each one, and presents them as actionable labels in the side panel;
(2)~\textbf{Actionable Guidance with Contextual Strategies}: when a user selects a right, \sys analyzes the current page's interactive elements in real time and provides step-by-step, highlighted guidance that adapts to the organization's stated mechanism---walking users through multi-page navigation paths, presenting direct links to privacy portals, or generating email templates---while tracking progress across page transitions and interface changess; and
(3)~\textbf{Transparent Evidence and Rights Education}: at any point during the guided process, users can request on-demand evidence grounded in the organization's own privacy policy, including verbatim policy excerpts and source links with alongside proactive educational notes that correct common misconceptions about the scope of each right, transforming mechanical task completion into informed action.
Through these mechanisms, \sys bridges the policy--practice divide by turning static legal disclosures into interactive, context-sensitive workflows that empower users to discover, exercise, and understand their data rights. 

To validate our design, we conducted a technical evaluation of \sys's rights extraction pipeline across 14 websites spanning multiple industries and popularity tiers, measuring precision, recall, and task completion rate against manually constructed ground truth.
We further conducted a qualitative user study ($N$=15) comparing \sys to a baseline condition, examining task completion rates, confidence levels, and rights understanding.
Our findings show that users who struggled to exercise privacy rights unaided consistently completed tasks with Privy's guidance, reporting increased confidence and discovering previously unknown rights and data practices. 


This paper makes the following contributions:
\begin{enumerate}
    \item The design and implementation of \sys, an LLM-empowered browser assistant that integrates policy analysis, mechanism-aware navigation guidance, and rights education into a unified system for DSR exercise on any website.
    \item A technical evaluation demonstrating \sys's ability to accurately extract organization-specific rights and exercising mechanisms from privacy policies on 14 diverse websites.
    \item A user study ($N$=15) providing empirical evidence that \sys improves task completion, user confidence, and rights understanding compared to unassisted DSR exercise.
    \item Design implications for building interactive privacy tools that bridge the policy--practice divide, informed by the challenges and opportunities surfaced through our evaluation.
\end{enumerate}

\section{Related Work}
\label{sec:related-work}

\subsection{The Policy–Practice Divide in Privacy Rights}

\begin{table*}[t]
\centering
\caption{Comparison of \sys with prior work across four capabilities identified in our review. \CIRCLE~= fully supported, \LEFTcircle~= partially supported, \Circle~= not supported.}
\label{tab:comparison}
\small
\begin{tabular}{lcccc}
\toprule
\textbf{System} & \textbf{Policy comprehension} & \textbf{Context-aware rights extraction} & \textbf{Interactive guidance} & \textbf{Transparency \& education} \\
\midrule
Nutrition Label~\cite{kelley2009nutrition}    & \CIRCLE     & \Circle     & \Circle     & \Circle     \\
Polisis~\cite{fawaz2017polisis}               & \CIRCLE     & \Circle     & \Circle     & \Circle     \\
PrivacyInjector~\cite{windl2022automating}    & \CIRCLE     & \Circle     & \Circle     & \Circle     \\
PRISMe~\cite{freiberger2025you}               & \CIRCLE     & \Circle     & \Circle     & \Circle     \\
Leschke et~al.~\cite{leschke2023streamlining} & \Circle     & \Circle     & \LEFTcircle & \Circle     \\
GPC~\cite{zimmeck2023usability}               & \Circle     & \Circle     & \Circle     & \Circle     \\
WebArena~\cite{zhou2024webarena}              & \Circle     & \Circle     & \LEFTcircle & \Circle     \\
\midrule
\textbf{\sys (Ours)}                          & \CIRCLE     & \CIRCLE     & \CIRCLE     & \CIRCLE     \\
\bottomrule
\end{tabular}
\end{table*}

Privacy regulations (e.g., GDPR, CCPA, PIPL) formally establish rights to access, deletion, and control, yet real-world exercise remains difficult~\cite{calzada2022citizens,zhou2024understanding,birrell2024sok,petelka2022generating,bowyer2022human,pins2022finding,zimmeck2023usability,borem2024data}.
Prior work characterizes this as a policy--practice divide, i.e., privacy rights are documented, but not reliably actionable in everyday interfaces.
Rather than reflecting a single breakdown point, this divide emerges across the full user journey from understanding a right, to locating the correct control, to confirming an outcome.
In this sense, legal compliance artifacts and interaction design artifacts jointly shape whether rights are practically exercisable.

This divide appears at two coupled layers.
At the \textit{comprehension} layer, users face long, legalistic policies with weak procedural guidance~\cite{mcdonald2008cost,kelley2009nutrition,schaub2015design,windl2022automating,amos2021policies,kaushik2021know}.
Prior studies and policy-analysis systems report recurring gaps between abstract disclosure and concrete actionability~\cite{bowyer2022human,petelka2022generating,fawaz2017polisis,pan2024new,freiberger2025you,chen2025clear,andow2019policylint,zhou2023policycomp,manandhar2022smarthome}.
At the \textit{interaction} layer, entry points are frequently hidden, fragmented, or manipulative~\cite{gray2018darkpatterns,mathur2019darkpatterns,pins2022finding,habib2020scavenger,nouwens2020dark,habib2022okay,boumasims2023cmp,lobel2024access}.
Compliance and enforcement studies further show mismatches between stated controls and actual behavior~\cite{degeling2019cookies,matte2020cookie,zimmeck2023usability,bollinger2022cookieblock,khandelwal2023cookieenforcer}.
Across this literature, users struggle from discovery through interpretation of outcomes~\cite{pins2022finding,bowyer2022human,petelka2022generating,borem2024data,habib2020scavenger, kaushik2021know}.
These gaps motivate systems that integrate policy interpretation, interface-grounded execution, and confidence-building feedback rather than optimizing only one layer.
It also suggests that incremental improvements at only one stage (e.g., better summaries without better routing) are unlikely to close the end-to-end usability gap.

\subsection{Privacy Policy Analysis and Comprehension Tools}

A substantial line of work improves policy comprehension via structured displays, NLP/ML parsing, machine-readable representations, and contextual explanation tools~\cite{kelley2009nutrition,schaub2015design,adjerid2013sleights,fawaz2017polisis,windl2022automating,pan2024new,freiberger2025you,chen2025clear,obar2020biggestlie,grunewald2023enabling}.
This line of work has further expanded to longitudinal and domain-specific audits that reveal policy drift, internal contradiction, overbroad claims, and coverage gaps~\cite{amos2021policies,andow2019policylint,zhou2023policycomp,manandhar2022smarthome}.

To date, these studies have established important foundations by improving readability, exposing semantic inconsistencies, and making policy content computationally tractable.
They also demonstrate that policy text can be operationalized for downstream tooling, including risk surfacing and context-aware explanation.
Several systems additionally show that presenting policy insights in situ (rather than in standalone documents) improves the likelihood that users engage with privacy content during real tasks.
At the same time, these benefits are typically bounded by the same limitation. That is, comprehension is improved without fully resolving procedural uncertainty about where and how to act on a specific site.

These systems clearly improve interpretability, but most remain informational: users may better understand policies while still lacking site-specific execution support for rights requests~\cite{fawaz2017polisis,windl2022automating,pan2024new,freiberger2025you,chen2025clear}.
In practice, users still need to map high-level policy statements onto concrete interface actions, which is where many rights workflows continue to fail.
This motivates approaches that bind policy extraction outputs to executable interaction pathways rather than presenting policy content as an endpoint.

\subsection{DSR Exercise Support and Guidance}

Plenty of complementary work has targeted DSR execution itself.
Studies of access/ opt-out workflows consistently report multi-step friction, unstable routing paths, and high failure in early discovery stages~\cite{pins2022finding,bowyer2022human,petelka2022generating,habib2020scavenger,borem2024data,leon2012johnny,habib2022okay,boumasims2023cmp,matte2020cookie,degeling2019cookies,nouwens2020dark}.
Automation approaches (e.g., scripted replay and protocol-level signals) reduce effort but can be brittle under UI changes or opaque from the user perspective~\cite{leschke2023streamlining,zimmeck2023usability,birrell2024sok,bollinger2022cookieblock,khandelwal2023cookieenforcer}.
Meanwhile, web-agent advances demonstrate strong page-grounded interaction, but generalist agents are not privacy-specialized by default~\cite{zhou2024webarena,deng2023mind2web,englehardt2016online}.

This body of work suggests a persistent tradeoff between generality and reliability.
Generic automation can be transferred across sites but often lacks organization-specific legal context, whereas handcrafted workflows are more concrete but less robust to interface drift.
Recent agentic interaction capabilities reduce part of this gap by adapting to live DOM structure, yet privacy rights semantics and compliance expectations still need explicit modeling.
Another recurring tension is visibility, i.e., low-friction automations can hide process details from users, whereas transparent, stepwise guidance can support trust and learning but may require additional interface design effort.

\subsection{User Confidence in Privacy Rights Exercise}

Even when tasks are completed, confidence often remains fragile~\cite{habib2020scavenger,ausloos2018shattering,politou2018forgetting,lobel2024access,birrell2024sok,adjerid2013sleights}.
Prior work points to two key facotrs.

\paragraph{The literacy gap.}
Users frequently hold incomplete mental models of data infrastructures, permissions, and legal scope~\cite{wash2010folk,lin2012expectation,almuhimedi2015location,felt2012android,kang2015mydata,turow2023consent,acquisti2015privacy,draper2019digital}.
As a result, many cannot confidently infer what rights cover, when to invoke them, or what consequences to expect.
This uncertainty can suppress action even when formal rights are available, because users cannot reliably predict either effort or benefit.
This is not simply an information-deficit problem. Several studies frame it as a rational response to opaque ecosystems in which consequences are difficult to observe directly.

\paragraph{The transparency gap.}
After users act, organizations often provide ambiguous confirmations or incomplete evidence, making outcomes difficult to verify~\cite{habib2020scavenger,ausloos2018shattering,politou2018forgetting,lobel2024access}.
This contributes to a persistent notice-versus-understanding gap documented across privacy communication research~\cite{obar2020biggestlie,schaub2015design}.
More broadly, policy comprehension tools and DSR automation tools tend to address different points in the workflow, leaving limited in-situ support for understanding why each step matters and what compliance outcomes should follow~\cite{fawaz2017polisis,windl2022automating,freiberger2025you,leschke2023streamlining,zimmeck2023usability}.
As such, the literature highlights that usability and trust are intertwined, suggesting that reducing clicks is insufficient if users cannot interpret or verify what happened.
This insight helps explain why systems that improve completion rates may still produce cautious or qualified user attitudes in post-task evaluations.


\begin{figure*}[!t]
  \centering
  \includegraphics[width=\textwidth]{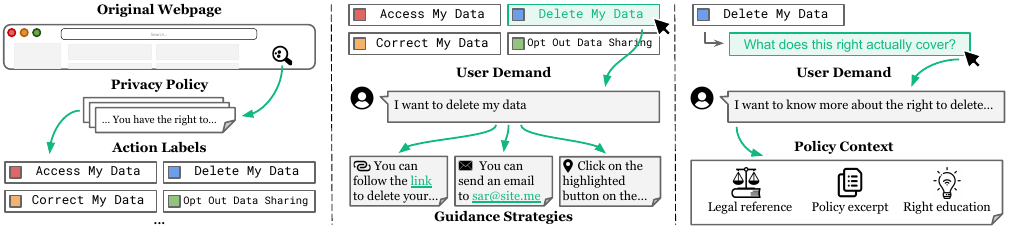} 
  \begin{subfigure}{0.3\textwidth}
    \caption{Policy Discovery \& Analysis}
  \end{subfigure}
  \hfill
  \begin{subfigure}{0.3\textwidth}
    \caption{Actionable Guidance}
  \end{subfigure}
  \hfill
  \begin{subfigure}{0.33\textwidth}
    \caption{Transparent Evidence and Literacy Support}
  \end{subfigure}
  \caption{Design Framework of \sys. (a) \sys automatically locates and analyzes a website's privacy policy, extracting applicable rights as action labels; (b) When a user selects a right, Privy adapts its guidance to the organization's stated mechanism: providing direct links, drafting emails, or highlighting UI elements for step-by-step navigation; (c) At any point, users can request policy context including a plain-language legal reference, the verbatim policy excerpt from the target website, and right education notes about the right being exercised.} \label{fig:framework} 
\end{figure*}

\section{System Design and Implementation}

\subsection{Overview}

This section presents the design of \sys, an interactive browser-based assistant that supports the general public in exercising their data subject rights (DSRs) across diverse organizations and websites. First we introduce the design goals in Section~\ref{sec:design-goals}. We then detail the realization for each design goal respectively. Lastly, we surface the implementation details of \sys in Section~\ref{sec:impl}

\subsection{Design Goals}
\label{sec:design-goals}

Deriving from the prior work and the policy--privacy divide, we propose \sys with the following design goals: 

\textit{[DG1] Contextualize Rights within Organizational Policies.} While privacy regulations grant individuals DSRs, organizations implement them through their own policies, specifying which rights are supported, how each can be exercised, and what conditions apply. For most users, identifying these rights requires reading lengthy and complicated policies that they often tend to skip~\cite{mcdonald2008cost}. Therefore, \sys should automate the extraction and analysis of the applicable rights from the policies and present them as concrete, site-specific actions to the users, bridging the gap between abstract legal entitlements and practical options.
    
\textit{[DG2] Provide Dynamic Guidance within Organizational Context.} Even when users know what right to exercise, the path to exercising it varies dramatically across organizations: some require navigating through nested settings menus, others provide a direct URL to a privacy portal, and others accept only email requests. These mechanisms are frequently buried in complex interfaces and even dark patterns. Thus, \sys should provide detailed and accurate guidance that adapts to each mechanism type.

\textit{[DG3] Amplify User Confidence through Transparent Evidence and Literacy Support.} Even after successfully completing a privacy action, users often remain uncertain due to the literacy gap~\cite{kang2015mydata,acquisti2015privacy} and transparency gap~\cite{habib2020scavenger,politou2018forgetting}. \sys should address these by providing on-demand evidence grounded in the organization's own policy to close the transparency gap, and proactive education about the legal scope of each right to close the literacy gap.

\subsection{Policy Discovery and Analysis}
\label{sec:policy-discovery}

To achieve DG1, \sys implements a pipeline that runs automatically when a user opens the side panel on any website (Figure~\ref{fig:framework}(a)). The pipeline begins with \textbf{policy discovery}: \sys locates the organization's privacy policy link on their domain, fetches the policy content, and extracts readable text. The extracted policy text is then sent to the backbone LLM for comprehensive analysis: identifying the applicable DSRs on the target website, extracting the relevant policy language for each right, and determining the concrete mechanism through which each right can be exercised.

From the user's perspective, once \sys has scanned and analyzed a website's privacy policy, several \textbf{Action Labels} appear in the side panel (Figure~\ref{fig:interface}(a)). By surfacing these options proactively, \sys alleviates the burden of identifying relevant actions, which is significantly beneficial for general users who may struggle to articulate their needs or even recognize which rights they hold. Unlike generic privacy tools that present a fixed menu of rights regardless of context~\cite{leschke2023streamlining,zimmeck2023usability}, \sys provides dynamically tailored options based on each organization's specific policy.

\begin{figure*}[!t]
  \centering
  \includegraphics[width=\textwidth]{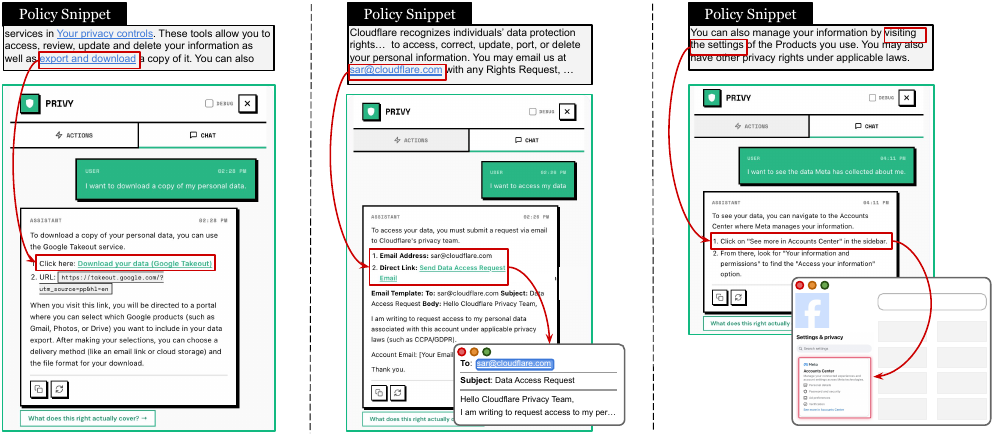} 
  \begin{subfigure}{0.28\linewidth}
    \caption{Strategy 1: Link}
  \end{subfigure}
  \hfill
  \begin{subfigure}{0.32\linewidth}
    \caption{Strategy 2: Email}
  \end{subfigure}
  \hfill
  \begin{subfigure}{0.35\linewidth}
    \caption{Strategy 3: Navigation}
  \end{subfigure}
  \caption{
    The three strategies that \sys will provide to the users for exercising their rights:
    (a) Link: \sys presents a direct URL to the organization's privacy portal and explains what to expect upon visiting;
    (b) Email: \sys generates a ready-to-use email template with pre-filled recipient, subject, and body.
    (c) Navigation: \sys highlights relevant UI elements on the page and provides step-by-step instructions across page transitions.
  } 
  \label{fig:strategies} 
\end{figure*}

\subsection{Actionable Guidance with Contextual Strategies}
\label{sec:actionable-guidance}

To accomplish DG2, \sys provides three mechanism-specific \textbf{guidance strategies}. When a user selects an action label, \sys reads the mechanism type stored within the policy analysis and adapts its guidance accordingly (Figure~\ref{fig:strategies}):

\textbf{Strategy 1: Link}
    Some organizations provide a direct URL to a privacy portal or request form (e.g., Google's Takeout page for data download; Figure~\ref{fig:strategies}(a)). \sys extracts the URL, presents it as a clickable link, and explains what the user should expect upon visiting.

\textbf{Strategy 2: Email}
    Some organizations require users to exercise a right via email (e.g., Cloudflare directs access requests to \textit{sar@cloudflare.com}; Figure~\ref{fig:strategies}(b)). \sys generates a ready-to-use email template with a pre-filled recipient address, subject line, and body text, ensuring that users can act on their rights without drafting a formal request from scratch.

\textbf{Strategy 3: Navigation}
    Many organizations provide only vague instructions, such as Meta's ``To exercise your rights, visit your settings for Facebook and Instagram'' (Figure~\ref{fig:strategies}(c)). For these cases, \sys explores the website, identifies the most promising path, and highlights UI elements step by step with visual overlays and concise instructions across pages.

\subsection{Transparent Evidence and Literacy Support}
\label{sec:literacy-support}

Unlike DG1 and DG2, which operate as a sequential pipeline, we design DG3 as an on-demand component: it is available throughout the guided process but only surfaces when users request it (Figure~\ref{fig:framework}(c)). Below each assistant response, a button of "What does this right actually cover?" invites users to learn more.
When activated, \sys draws on privacy policy of the website and generates a \textbf{policy context}---a structured explanation comprising three components:

\textit{Legal reference}: a concise explanation of what the right means under the applicable regulation (e.g., CCPA), translating legal language into plain terms that users without a law background can understand.

\textit{Policy excerpt}: the verbatim text from the organization's own privacy policy that establishes the right, with a link to the source URL so users can verify it independently.

\textit{Rights education}: a brief note that addresses a common misconception about the right being exercised.
For example, when a user opts out of data sales, \sys might display: ``\textit{Most people think this means: Unsubscribing from marketing emails. Actually: The company must stop selling your data to third parties AND stop sharing it for targeted advertising---even if no money changes hands.}''

Together, these three components create a transparent evidence chain---from the user's action to the organization's policy to the legal guarantee---enabling users to audit \sys's recommendations rather than accepting them on faith.

\subsection{Implementation}
\label{sec:impl}

\sys is implemented as a Chrome Manifest V3 browser extension comprising approximately 3{,}500 lines of vanilla JavaScript with no external framework dependencies.
Page understanding relies on the Chrome DevTools Protocol to capture the full accessibility tree of the active tab, which \sys uses to identify interactive elements and ground the LLM's guidance in the live page structure. We use Google Gemini-3.0-Flash as the default backbone LLM, with adapter support for Anthropic Claude-Sonnet-4 and OpenAI GPT-4o.
Prompts, few-shot examples, and sample outputs are available in the supplementary materials.

\section{Technical Evaluation}
\label{sec:evaluation}

We conducted a technical evaluation to validate \sys's two core LLM pipelines before deploying the system with human participants: (1)~rights extraction from privacy policies, and (2)~end-to-end workflow guidance for completing privacy tasks.

We tested \sys on 14 websites across seven industries: social media (Facebook, Reddit, X, LinkedIn), e-commerce (Amazon, Walmart, Target), media (Netflix, Spotify, NYTimes), SaaS (Google, Zoom), travel (Airbnb), and health (WebMD).
For each site, a researcher read the full privacy policy, identified all CCPA/CPRA rights, and recorded the mechanism and action value for each right.
This produced a ground-truth corpus of 185 rights across the 14 sites.

\subsection{Privacy Rights Extraction Performance}

We measured rights extraction using standard precision and recall.
Precision captures whether users can trust the action labels \sys presents. That is, whether each action label presented to users corresponds to a genuine policy right.
Recall captures whether \sys misses important rights.

\begin{table}[t]
\centering
\caption{Rights extraction performance across 14 websites.}
\label{tab:rights-extraction}
\small
\begin{tabular}{lcccc}
\toprule
\textbf{Site} & \textbf{action labels} & \textbf{GT Rights} & \textbf{Precision} & \textbf{Recall} \\
\midrule
Spotify       & 11 & 13 & 1.000 & 0.846 \\
Airbnb        &  4 &  9 & 0.800 & 0.444 \\
Amazon        & 10 & 11 & 1.000 & 1.000 \\
Facebook      & 11 & 11 & 1.000 & 0.909 \\
Google        & 10 & 13 & 1.000 & 0.769 \\
LinkedIn      & 13 & 16 & 1.000 & 0.812 \\
Netflix       & 12 & 15 & 1.000 & 0.867 \\
NYTimes       & 11 & 17 & 1.000 & 0.824 \\
Reddit        & 11 & 14 & 0.909 & 0.714 \\
Target        & 11 & 16 & 1.000 & 0.750 \\
Walmart       & 11 & 12 & 1.000 & 1.000 \\
WebMD         &  9 & 14 & 1.000 & 0.786 \\
Zoom          & 10 & 12 & 1.000 & 0.833 \\
X (Twitter)   & 11 & 12 & 1.000 & 0.833 \\
\midrule
\textbf{Mean} & \textbf{10.4} & \textbf{13.2} & \textbf{0.979} & \textbf{0.813} \\
\bottomrule
\end{tabular}
\end{table}

Table~\ref{tab:rights-extraction} presents the per-site results.
Precision was near-perfect (0.979). Only 2 of 145 extracted action labels did not match a ground-truth right, and no hallucinated rights were observed on any site.
Every site produced at least three correctly extracted right, meaning \sys was immediately useful on all 14 sites.

Recall was 0.813 (macro F1 = 0.885), meaning \sys identified the majority of rights stated in each policy.
The missed rights were predominantly ones with limited practical impact: declarative rights that users cannot act on (e.g., ``Right to Non-Discrimination''), granular platform-specific controls (e.g., Reddit's cookie and location settings), and jurisdiction-specific rights outside the CCPA scope (e.g., WebMD's EEA consent withdrawal).
Only a small number of core, actionable CCPA rights were genuinely missed, most notably Facebook's right to correction.
Airbnb was a clear outlier with a recall of 0.444, likely due to its heavily cross-referenced, multi-layered policy structure. Excluding Airbnb, mean recall rises to 0.841.

\subsection{End-to-End Workflow Performance}

An evaluator followed \sys's instructions on each site for up to four task types: data access, deletion, opt-out, and correction.
Each task was coded as \textit{success} (reached the expected end state), \textit{partial} (reached the correct area but could not complete the final step), or \textit{failure} (never reached the correct area).
We also recorded the number of steps from first prompt to completion.

\begin{table}[t]
\centering
\caption{End-to-end workflow performance by task type across 14 websites.}
\label{tab:workflow}
\small
\begin{tabular}{lccccc}
\toprule
\textbf{Task Type} & \textbf{Total} & \textbf{Success} & \textbf{Partial} & \textbf{Failure} & \textbf{Mean Steps} \\
\midrule
Access     & 14 & 14 & 0 & 0 & 4.1 \\
Delete     & 14 & 13 & 1 & 0 & 3.0 \\
Opt-out    & 13 & 13 & 0 & 0 & 2.4 \\
Correction & 13 & 12 & 1 & 0 & 3.1 \\
\midrule
\textbf{All} & \textbf{54} & \textbf{52} & \textbf{2} & \textbf{0} & \textbf{3.2} \\
\bottomrule
\end{tabular}
\end{table}

\sys completed 96.3\% of tasks successfully (52/54) with zero complete failures (Table~\ref{tab:workflow}).
The mean number of steps was 3.2.
The two partial outcomes both involved deeply nested interfaces.
On Facebook, the delete task entered a stuck loop: \sys repeatedly cycled between two sub-menus in Meta's Accounts Center without finding the deletion option.
On Netflix, the correction task hit a redirect loop: \sys suggested a ``Manage privacy'' link that redirected to the homepage, and repeated the same suggestion upon return.

Opt-out tasks were the easiest (2.4 steps, 100\% success), as they typically require a single toggle or short navigation path.
Access tasks took the most steps (4.1) but all succeeded.
Delete tasks were efficient (3.0 steps) with one partial outcome.
Correction had one partial outcome as well—most sites lack a dedicated correction page, making it harder for the LLM to locate the right interface element.

We also observed a clear efficiency difference by mechanism type.
When the policy provided a direct link (e.g., Google Takeout, Netflix's data request portal), tasks completed in 1--3 steps.
Navigation-based tasks, which require highlighting elements across multiple page transitions, averaged 3--6 steps.
This validates the pipeline design: extracting the mechanism before initiating guidance allows \sys to take the most efficient path.

\subsection{Summary}

\sys's rights extraction achieves high precision (0.979) with no hallucinated rights and sufficient recall (0.813) to surface the core actionable rights on every site tested.
The workflow guidance completes 96.3\% of tasks in an average of 3.2 steps.
These results confirm that the system is suitable for deployment with end users, which we evaluate in the following section.

\section{User Study}
\label{sec:user-study}

To understand how \sys affects real users' ability to exercise their privacy rights, we conducted a qualitative user study with 15 participants.
The study was designed to address three research questions:

\textbf{RQ1:} How effectively does \sys bridge the policy--practice divide in exercising data subject rights?

\textbf{RQ2:} How does using \sys influence users' awareness of and confidence in their data subject rights?

\textbf{RQ3:} Do users find \sys usable and useful?

\subsection{Participants}

We recruited 15 participants through Prolific.
Participants were required to complete a pre-screening survey that collected demographic information (age, gender, education), technical proficiency (browser usage frequency, computer science background), and prior experience with data subject rights.
We sought to include participants with diverse levels of technical background and privacy awareness.
Each participant was compensated \$20 USD for a session lasting approximately 40 minutes.

\subsection{Study Design}

Each session was conducted remotely via Zoom and consisted of three phases. With participants' consent, all sessions were recorded for subsequent analysis.

\textbf{Phase 1: Establishing Baseline Privacy Awareness and Behavior} 
    The session began with a semi-structured interview exploring participants' relationship with online privacy: how they manage personal data, what challenges they face, and what they understand about their legal rights. We then probed participants' prior experience with specific DSRs---whether they had heard of the rights to access, delete, correct, or opt out of data sharing, and whether they had ever attempted to exercise any of these rights on a website. To establish a behavioral baseline, we asked each participant to attempt a specific privacy task on a real website without \sys. For example, "Try to access your data on YouTube" or "Try to opt out of data sharing on Amazon". This unaided attempt served as each participant's own control, allowing us to observe firsthand the barriers they encountered and the strategies they tried before introducing \sys.

\textbf{Phase 2: Exercising Rights with \sys}
    Participants were then introduced to \sys through a brief walkthrough of the side panel interface. They were asked to complete up to four privacy tasks on real websites using \sys: deleting data on YouTube, accessing data on Google, opting out of data sharing on Amazon, and a free-play period where they could explore \sys on any website of their choosing. Participants were instructed to think aloud throughout the tasks, verbalizing their expectations, reactions, and any confusion as they interacted with the system. The researcher observed but did not intervene unless the participant was stuck for an extended period.

\textbf{Phase 3: Reflecting on the Experience}
    After completing the tasks, participants filled out the System Usability Scale (SUS) questionnaire~\cite{brooke1996sus}. This was followed by a semi-structured interview in which participants reflected on their overall experience with \sys---what they liked, what challenges they encountered, and whether anything surprised them during the process. We specifically asked whether they had discovered any rights they did not previously know they had, and how their confidence in exercising privacy rights had changed. Participants rated their post-session confidence on a 5-point scale and were asked to explain their rating, enabling us to capture both the direction and reasoning behind any confidence shift.

\subsection{Data Analysis}

All sessions were recorded and transcribed. For qualitative analysis, we followed established open coding procedures~\cite{strauss1990basics}. Two members of the research team independently coded 20\% of the transcripts and generated an initial set of codes. The two researchers then discussed disagreements and reconciled differences to produce a cohesive codebook. Using this codebook, we conducted a thematic analysis across all transcripts to identify recurring patterns relevant to the three research questions.

SUS scores were computed following the standard scoring procedure and are reported descriptively (mean, standard deviation). Confidence ratings from the closing interview are reported descriptively alongside participants' qualitative explanations of their ratings.

\section{Findings}

\begin{figure}[t]
  \centering
  \includegraphics[width=\linewidth]{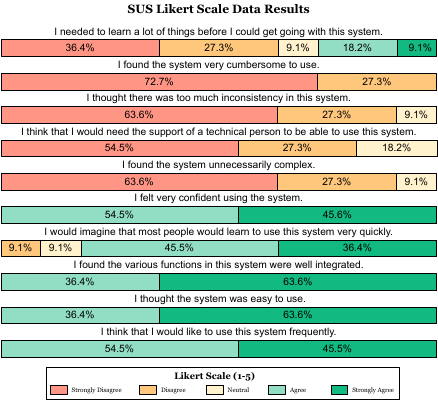} 
  \caption{Result of SUS quesionaire. For the negative statements (first fve), lower ratings are better. For the positive statements (bottom fve), higher is better}
  \label{fig:sus} 
\end{figure}

Our user study reveals three key findings.
First, participants across all levels of privacy awareness were struggling to exercise their rights unaided, but consistently succeeded when guided by \sys (\textbf{RQ1}).
Second, using \sys led participants to discover rights and data practices they had not previously known about, and participants reported increased confidence in exercising their rights (\textbf{RQ2}).
Third, participants rated \sys as highly usable and expressed strong interest in continued access (\textbf{RQ3}).

\subsection{RQ1: Users struggle to exercise rights unaided but complete tasks with \sys's guided navigation}

To address RQ1, we organize our analysis around three themes that characterize participants' navigation experiences, i.e., difficulty locating rights pathways, the role of guided visibility in supporting action, and how successful navigation reshaped participants' interpretations of platform design.

\textbf{Rights pathways are hard to locate, regardless of baseline awareness.}
Across interviews, participants described rights exercise as primarily a navigation problem rather than a lack of interest.
When asked to attempt at least one DSR task on YouTube without assistance, participants across the full awareness spectrum failed to complete the tasks. These include participants who had never considered privacy (e.g., P3: ``I didn't think about things like that'') to those with formal privacy training (P11) or highly restrictive personal settings (P7: ``I've literally turned everything off'').
Most followed a familiar but unproductive path (profile $\rightarrow$ settings $\rightarrow$ "privacy"). As P3 summarized: ``Profile... Settings... Privacy... No idea.''
Even participants with stronger baseline knowledge did not reach completion. P8 correctly reasoned that ``YouTube's owned by Google, I would have to go to my Google account,'' and P11 had previously seen Google Takeout, but neither could locate the exact rights-exercise path starting from YouTube.
P2 described the process as ``a rabbit hole of clicking here and going there, reading this, trying to figure out... it was not a feat that most people would do.''

\textbf{Guided visibility supports orientation, completion, and learning.}
Participants characterized \sys as making hidden workflows legible by showing where to go next in context.
The most frequently praised feature was visual highlighting of interface elements: seven participants said it prevented disorientation and clarified next actions.
P1 noted, ``It highlighted where you should go next so that way you didn't get lost.'' P5 emphasized efficiency, ``For me to do that would take ages. And so the fact that it takes just a few minutes, I really like.''

Participants also valued that \sys guides rather than automates. As P1 put it, ``I like that it doesn't just do it for you, that it tells you how to do it...because now I know where to go next time, even without it.''
This interpretive support corresponded to concrete in-session outcomes: all participants completed at least one privacy task, and four participants (P7, P8, P11, P12) used \sys to verify that they had already opted out on certain platforms.

\textbf{Successful navigation reveals perceived organizational opacity.}
Using \sys not only helped participants complete tasks but also altered how they interpreted the platform design.
Eleven of 15 participants recognized, often for the first time, that privacy controls appeared obscured by complex navigation, jargon-heavy policy language, and unintuitive naming.
P2 observed, ``They're just burying all this information and it's just not full disclosure. To me, it's just... it's wrong.'' P11 noted naming issues: ``They use a weird title called Google Takeout instead of, like, Google your Google Data and Privacy.''
Two participants (P9, P10) explicitly described these patterns as deceptive; P9 said the placement of Google Takeout ``does seem deceptive, or probably intentionally deceptive in order to keep that information.''

Taken together, these themes indicate that \sys's value in RQ1 is not only task success but also procedural legibility: participants could find, interpret, and verify rights-related actions that were otherwise difficult to execute.

\subsection{RQ2: Using \sys increases rights awareness and self-reported confidence}

\subsubsection{Three distinct forms of awareness gains}

\textbf{Discovering that rights and controls exist.} 
For some participants, the central shift was simply learning that specific controls were available to them.
P1 said, ``I didn't know that you could actually turn off your ads. Like I'm going to go do that when we're done here.''
P10 echoed this, ``I didn't know exactly what YouTube was tracking on me, and I did not know I had the ability to turn that off.''
P3, who had never considered online privacy before the study, expressed a more foundational realization, ``I'm surprised that I have privacy using website. Because I never know the website can have like... huge of my information.''
Seven of 15 participants reported this kind of discovery.

\textbf{Discovering how to exercise known rights.}
Other participants entered the study with general awareness that rights exist, but lacked procedural knowledge for enacting them.
P7 captured this gap, ``I was aware of some of the laws that's been passed where we can kind of control our data, but I really had no idea about how to do that.''
Five of 15 participants described this shift from abstract awareness to actionable understanding.

\textbf{Discovering hidden data-sharing practices.}
Participants also reported learning new information about how organizations use and share their data through \sys's transparent evidence and literacy support.
P7 was surprised to learn about cross-platform data selling, ``I did not know that Facebook was selling my information to Amazon. That I did not know.''
P10 described a similar realization involving financial services: after receiving a targeted call from Edward Jones based on stock holdings at Fidelity, P10 inferred that ``they were sharing information about what stocks I held.''
Eight participants reported this type of discovery, which broadened their understanding of data flows beyond the platforms they directly use.

These themes show that \sys supported layered awareness gains from recognizing rights to learning procedural steps, and eventually, to understanding downstream data practices.

\subsubsection{Confidence increases, with residual skepticism.}
All 15 participants reported increased confidence in exercising privacy rights after using \sys, with self-reported ratings ranging from 4 to 5 on a 5-point scale ($Mdn$ = 4).
We organized their feedback into the following three themes.

\textbf{Actionable controls increase perceived agency.}
Participants linked confidence gains to seeing concrete, available actions. 
P1 explained, ``Just that there are options out there to protect your privacy, because I know it's a big issue.''
P2, who rated 5 out of 5, said, ``It does give me a sense of being able to control my own data... it's a huge step forward for consumer rights.''
P10 similarly expressed intent to act beyond the study, ``I feel like using that tool, there's a lot of websites I'd like to go through to see what rights I have, and scrub my data.''
These examples suggest that confidence was grounded not only in successful task completion, but in a stronger sense of personal agency and readiness to take continued privacy action beyond the study context.

\textbf{Plain language explanations support confidence across experience levels.}
Participants also credited confidence gains to explanations that translate legal and technical languages into understandable terms. 
For example, P1 noted, ``There is ways to learn about it now... I get an explanation that isn't just technical speak.''
P7, who rated 4 out of 5 despite being relatively tech-savvy, said, ``It's giving me information that I would have never found on my own. I'm relatively tech savvy, more so probably than the average person. And I had no idea about that.''
These responses indicate that interpretability itself functioned as a confidence mechanism, helping participants with different prior expertise translate abstract rights language into actionable understanding.

\textbf{Confidence is tempered by institutional and functional uncertainty.}
Even with increased confidence, some participants remained skeptical about whether organizations would honor their requests.
Four participants raised this concern. P2 asked, ``Do they really have to do that? They really do it? That's my question.''
When told that companies are legally obligated to comply, P2 responded, ``Yeah, I know, but who's going to do that?''
Two participants (P9, P10) also worried about whether exercising a right through \sys might unintentionally disable needed functionality on the website.

 To summarize, these results suggest that \sys can raise confidence substantially, but sustained trust also depends on participants' beliefs about organizational compliance and downstream consequences.

\subsection{RQ3: Users rate \sys as highly usable and express demand for continued access}

\textbf{SUS results.}
Participants rated \sys's usability using the System Usability Scale~\cite{brooke1996sus}.
The mean SUS score was 85.2 ($SD$ = 10.3).
Figure~\ref{fig:sus} shows the per-item Likert distributions.
Specifically, no participant rated any positive item (ease of use, confidence, integration, learnability, frequency of use) below ``Somewhat Agree.''
No participant rated any negative item (complexity, cumbersomeness, inconsistency, need for technical support) above ``Neither Agree nor Disagree.''
The weakest item was Q10 (``I needed to learn a lot of things before I could get going with this system''), where 27.3\% of participants somewhat or strongly agreed, consistent with the qualitative feedback about onboarding discussed below.

\textbf{Perceived value and demand for access.}
Nine of 15 participants either asked when \sys would be publicly available, expressed intent to keep using it after the study, or described it as broadly needed.
P1 asked, ``Is it going to be available on the Google Chrome store at some point?''
P2 asked, ``So when can I have access to it?'' and added, ``It sounds just wonderful and like something that we all need at this time in the technological world.''
P7 stated, ``I'm not going to uninstall it. I might be playing with this a bit more once we're done with the call.''
P3, who had never previously considered online privacy, said, ``I think this tool is really convenient and useful. I think after a short learn, I can handle it and I will use it like every day to check my privacy in every website.''
This endorsement spanned the full range of participants, from those with no prior privacy knowledge to those with professional training.

\section{Discussion}

\subsection{Interpreting the Combined Evidence}
Taken together, the technical evaluation and user study suggest that the main bottleneck in exercising privacy rights is not the absence of legal rights, but the operational work required to translate formal rights into concrete, service-specific actions.
Our technical results show that this translation can be made reliable for most workflows when policy parsing and interface-grounded guidance are coupled, while the user study shows that participants who struggled unaided often completed the same tasks with targeted support.
This convergence matters because prior work has often documented these failures separately, i.e., policy-comprehension studies emphasize legal-language opacity~\cite{bowyer2022human,petelka2022generating}, while DSR-usability studies emphasize fragmented, failure-prone interface journeys~\cite{pins2022finding,zimmeck2023usability}.
Our results connect these strands by showing that the policy--interface gap is a single interaction problem experienced end-to-end by users.

More specifically, our findings extend prior accounts of ``digital resignation'' by suggesting that resignation is not only attitudinal, but also infrastructural: users disengage when systems demand high-effort interpretation across policies, settings hierarchies, and cross-brand account boundaries~\cite{draper2019digital}.
At the same time, participants' confidence gains and use of optional explanations indicate that successful task completion can become a practical entry point for privacy literacy, complementing prior work that treats comprehension and action as separate challenges.

Building on this synthesis, we next discuss why these failures persist at a system level and then derive concrete design implications.

\subsection{Why Existing DSR Ecosystems Fail Structurally}
Our findings suggest that many rights-exercise failures are structural rather than purely individual.
Across cases, friction emerged when controls were split across brands, domains, and account systems that users did not perceive as one privacy surface, and when entry points were inconsistently named or deeply nested.
The partial failures in our technical evaluation occurred in these cross-service transitions, reinforcing earlier evidence that rights pathways break at organizational boundaries even when formal mechanisms exist~\cite{leschke2023streamlining}.
Under these conditions, even motivated users with moderate privacy knowledge face high coordination costs and uncertain task boundaries.
This implies that improving DSR usability requires system-level alignment of architecture, labeling, and ownership boundaries, not only better user training or individual effort.

\subsection{Design Implications}
Our findings show that improving privacy outcomes requires coordinated changes at three layers: policy communication, interface architecture, and assistance-tool design.
Based on these results, we organize implications from foundational system structure to longer-term user support.

\textbf{Privacy Controls Should Be Unified Within Each Service, Not Fragmented Across Organizations.}
RQ1 indicates that rights workflows fail when controls are distributed across brands, domains, and account systems that users do not perceive as a single privacy surface.
Nearly all participants identified this fragmentation as confusing, and the two partial failures in our technical evaluation also occurred in deeply nested, cross-service interfaces.
For example, YouTube data export requires routing to Google Takeout, while Facebook settings are embedded in Meta's Accounts Center.
We recommend that each service expose stable, discoverable privacy entry points within its own interface, even when backend ownership is shared.
At a minimum, service-level entry points should provide explicit routing, progress cues, and clear confirmation states so users can understand where they are in the request lifecycle.
In short, privacy controls should be at least as visible and navigable as the data collection pathways they govern.

\textbf{Privacy Policies Should Front-Load Actionable Information in Plain Language.}
Our results suggest that policies should be restructured around user tasks, not only legal categories.
DG1 shows that the information users need (e.g., which right applies, where to submit a request, and what outcome to expect) is often present but buried within legal disclaimers and domain-specific language.
Participants repeatedly described this as a reason they had not attempted rights exercise previously.
We therefore recommend front-loading user-facing policy content in plain language, with concrete action cues (e.g., ``To opt out, visit [URL]'').
Importantly, this does not require removing legal precision; rather, policy documents can separate a concise action layer for users from a detailed compliance layer for legal review.
This layered structure can reduce comprehension burden while preserving regulatory fidelity.

\textbf{Designs Should Support Diverse Privacy Starting Points.}
A consistent result across RQ1 and RQ2 is that participants entered with highly variable prior knowledge, yet many encountered similar workflow barriers without guidance.
This suggests that effective privacy systems should not assume either novice ignorance or expert competence.
Instead, they should provide adaptive support that scales with user needs: clear default guidance for first-time users, optional detail layers for advanced users, and stable terminology that remains consistent across contexts.
Designing for heterogeneous starting points can improve accessibility without oversimplifying rights information.

\textbf{Task Completion Creates a Learning Moment for Privacy Education.}
RQ2 highlights a recurring tension between efficiency and understanding.
Participants valued completing tasks quickly, but confidence gains were strongest when actions were paired with plain-language explanations and contextual evidence.
DG3's on-demand design addressed this tension by allowing task-focused users to proceed without interruption while still supporting deeper learning when requested.
The majority of participants used the ``learn more'' feature at least once, suggesting that action moments can become high-receptivity moments for privacy education.
Future tools should therefore implement progressive disclosure: an action-first flow with optional, context-linked explanations of rights scope, implications, and expected outcomes.
This design can support immediate task success without sacrificing longer-term privacy literacy.

\textbf{Trust Must Be Designed as a First-Class Feature.}
RQ2 and RQ3 jointly show that usability gains alone are insufficient for sustained adoption if users remain uncertain about organizational compliance or the assistant tool's own data practices.
Even participants who reported higher confidence questioned whether companies would honor requests and whether using \sys could introduce new risks.
At the same time, learning that \sys collects no user data was treated as a major positive differentiator.
These findings suggest that privacy assistants should treat trust as a core product surface: clearly disclose what data is collected, provide verifiable logs or evidence of completed actions, and communicate limits of what automation can guarantee.
Designing for verifiability and transparency can convert short-term usability into durable trust and continued use.

\textbf{Privacy Support Should Extend Beyond One-Off Task Completion.}
RQ3 indicates demand for continued access, suggesting that users view privacy exercise as an ongoing practice rather than a single transaction.
This has an important design implication: tools should support longitudinal privacy management, not only one-time completion.
In practice, this could include lightweight history views of completed actions, reminders to revisit key settings, and follow-up checks when platforms change interfaces or policies.
Such continuity features can help users maintain control over time and reduce the need to relearn fragmented workflows from scratch.

\subsection{Cross-Context Implications and Trade-offs}
To situate these implications beyond our immediate setting, we clarify what may transfer across contexts and what tensions remain unresolved.

\textbf{Generalizability and Transferability.}
Several patterns in our results are likely to transfer across regulatory settings, including the need for actionable policy language, stable navigation pathways, and clear post-action confirmation.
These needs are consistent with prior evidence from GDPR- and CCPA-related workflows, and are plausibly relevant to PIPL contexts where users still face policy-to-interface translation burdens despite different legal framing.
At the same time, some aspects are context-dependent: terminology, identity-verification requirements, platform governance models, and baseline institutional trust may vary by region and service type.
We therefore view our contribution as a transferable interaction framework with jurisdiction- and platform-specific implementation details.

\textbf{Design Tensions and Trade-offs.}
Our study also highlights recurring trade-offs in privacy-assistance design.
Greater automation can reduce effort but may reduce user agency when decision logic is opaque; faster flows can increase completion but weaken comprehension if explanatory content is minimized.
Similarly, detailed guidance can improve immediate success yet increase over-reliance, while stronger transparency can build trust but raise cognitive load.
Future tools should therefore support adjustable guidance depth so users can move between efficient completion and deeper understanding based on their goals and experience.

\subsection{Limitations}
Our findings should be interpreted within several limitations.
First, the study evaluates rights exercise in a bounded set of high-traffic services and task types, so performance may differ for smaller organizations, less standardized workflows, or rapidly changing interface states.
Second, the user study captures short-horizon task execution and immediate reflections; it does not establish whether confidence gains persist over time or whether users continue exercising rights without support.
Third, while \sys improves pathway discovery and completion, it cannot guarantee downstream organizational compliance, response quality, or the completeness of fulfilled requests.

\section{Conclusion}

In this paper, we present \sys, an interactive LLM-based browser assistant that bridges the policy--practice divide in privacy rights exercise. Our study demonstrates that \sys effectively supports rights exercise by providing systematic, interactive, transparent, and context-aware guidance, enabling users to not only accomplish rights-related actions that are difficult to complete unaided, but also develop greater interest and confidence in privacy protection. 
We discuss design implications based on our results.

\section*{Acknowledgement}
    We used ChatGPT for proofreading.

\bibliographystyle{ACM-Reference-Format}
\bibliography{ref}

\appendix


\section{Codebook}
\label{appendix:codebook}

\begin{enumerate}[leftmargin=14pt]
    \item Baseline Privacy Attitudes
    \begin{enumerate}[label=(\alph*), nosep, leftmargin=12pt]
        \item Skipping privacy policies
        \item Privacy as convenience tradeoff
        \item Proactive privacy management
        \item Behavioral privacy strategies
        \item Privacy settings feel ineffective
        \item Privacy indifference
        \item Assumed corporate benevolence
    \end{enumerate}

    \item Barriers to Privacy Rights Exercise
    \begin{enumerate}[label=(\alph*), nosep, leftmargin=12pt]
        \item No DSR awareness
        \item Surface-level rights awareness
        \item No legal grounding
        \item Time and effort barrier
        \item Digital resignation
        \item Policy inaccessibility
        \item Public helplessness
        \item Wanting a tool
        \item Fear of breaking functionality
    \end{enumerate}

    \item Baseline Task Performance
    \begin{enumerate}[label=(\alph*), nosep, leftmargin=12pt]
        \item Baseline task failure
        \item Baseline task partial success
        \item Rabbit-hole navigation
        \item Frustration from complexity
    \end{enumerate}

    \item Experience with \sys{} --- Guidance
    \begin{enumerate}[label=(\alph*), nosep, leftmargin=12pt]
        \item Highlighting as wayfinding
        \item Valuing step-by-step guidance
        \item Streamlined vs.\ manual
        \item Experiencing task completion
        \item Confirming prior actions
        \item Valuing guided recovery
    \end{enumerate}

    \item Experience with \sys{} --- Education \& Evidence
    \begin{enumerate}[label=(\alph*), nosep, leftmargin=12pt]
        \item Discovering new rights
        \item Discovering how to exercise rights
        \item Data-practice surprise
        \item Organizational opacity recognized
        \item Education impact
        \item Expanding threat model
        \item Intentional deception recognized
    \end{enumerate}

    \item Confidence Shift
    \begin{enumerate}[label=(\alph*), nosep, leftmargin=12pt]
        \item Confidence increase
        \item Having options as empowerment
        \item Accessible explanations build confidence
        \item Compliance skepticism
        \item Recognizing official request channels
    \end{enumerate}

    \item Usability and Value
    \begin{enumerate}[label=(\alph*), nosep, leftmargin=12pt]
        \item Jargon translation
        \item Preferring guidance over automation
        \item Navigation needs polish
        \item Discoverability issue
        \item Desire for availability
        \item Seeing tool as broadly needed
        \item Task focus over learning
        \item Privacy concern about tool
    \end{enumerate}

    \item Context and Background
    \begin{enumerate}[label=(\alph*), nosep, leftmargin=12pt]
        \item Personal breach experience
        \item Generational privacy shift
        \item Retroactive regret
        \item AI profiling fear
        \item Platform distrust
        \item Mobile access barrier
    \end{enumerate}
\end{enumerate}

\section{LLM Prompts}
\label{appendix:prompts}

\begin{mdframed}[linewidth=0.5pt, innertopmargin=10pt, innerbottommargin=10pt, innerleftmargin=10pt, innerrightmargin=10pt, skipabove=10pt, skipbelow=8pt]
\small
\textbf{Prompt for privacy rights extraction (DG1)}

You are a privacy compliance analyst. Given a website's privacy policy excerpt, identify the consumer privacy rights it offers (especially CCPA/CPRA rights). Respond only with JSON in this exact shape:

\{``rights'': [\{\\
\quad ``id'': ``...'', ``label'': ``...'', ``prompt'': ``...'',\\
\quad ``excerpt'': ``...'',\\
\quad ``mechanism'': ``email'' $|$ ``link'' $|$ ``navigation'' $|$ ``form'',\\
\quad ``action\_value'': ``...'',\\
\quad ``keywords'': [...]\}]\}

Extract ALL consumer privacy rights stated in the policy. Include every distinct right that has its own mechanism or action.

\emph{mechanism}: How the user exercises the right (email, link, navigation, or form).

\emph{action\_value}: The specific email address, URL, or navigation path extracted from the text.

If the policy does not describe any rights, return \{``rights'': []\}.
\end{mdframed}

\begin{mdframed}[linewidth=0.5pt, innertopmargin=10pt, innerbottommargin=10pt, innerleftmargin=10pt, innerrightmargin=10pt, skipabove=10pt, skipbelow=8pt]
\small
\textbf{Prompt for step-by-step guidance (DG2)}

You are Privy, an expert privacy assistant helping users with CCPA rights.

1.~\textbf{Goal}: Help users complete privacy actions based on the provided page accessibility tree.

2.~\textbf{Context}: You will receive a JSON accessibility tree. Each node may have: \emph{role}, \emph{name}, \emph{privyId} (for interactive elements), \emph{children}, and state properties (\emph{disabled}, \emph{expanded}, \emph{checked}).

3.~\textbf{Response style}: One-sentence summary followed by numbered quick steps. Refer to UI elements by their accessible name.

4.~\textbf{Response structure}: Every response must contain three blocks:

[REASONING] Private chain-of-thought (not shown to user). [/REASONING]

[RESPONSE] User-facing instructions. [/RESPONSE]

[MACHINE\_OUTPUT]\\
\quad \{``highlights'': [\{``label'': ``...'', ``id'': ``<privyId>''\}]\}\\
{[/MACHINE\_OUTPUT]}
\end{mdframed}

\begin{mdframed}[linewidth=0.5pt, innertopmargin=10pt, innerbottommargin=10pt, innerleftmargin=10pt, innerrightmargin=10pt, skipabove=10pt, skipbelow=8pt]
\small
\textbf{Prompt for policy link selection (DG1 discovery)}

You are a privacy policy link selector. Given a list of links found on a webpage, select the SINGLE link most likely to be the site's main privacy policy page.

Prioritize: (1)~links named ``Privacy Policy'' or ``Privacy Notice''; (2)~links in footer or navigation regions; (3)~URLs containing /privacy; (4)~same-origin links.

De-prioritize: cookie settings, terms of service, third-party framework links, ``Do Not Sell'' pages.

Respond with JSON:\\
\{``selectedUrl'': ``...'' $|$ null,\\
\quad ``confidence'': ``high'' $|$ ``medium'' $|$ ``low'',\\
\quad ``reason'': ``...''\}
\end{mdframed}

\end{document}